# Nonequilibrium Energetics of Molecular Motor Kinesin

Takayuki Ariga[1,2*], Michio Tomishige[3], Daisuke Mizuno[2]

[1]*Graduate School of Medicine, Yamaguchi University, Yamaguchi 755-8505, Japan*
[2]*Department of Physics, Kyushu University, Fukuoka 819-0395, Japan*
[3] *Department of Physics and Mathematics, Aoyama Gakuin University, Kanagawa 252-5258, Japan*

(Dated: 10 July 2018)

Nonequilibrium energetics of single molecule translational motor kinesin was investigated by measuring heat dissipation from the violation of the fluctuation-response relation of a probe attached to the motor using optical tweezers. The sum of the dissipation and work did not amount to the input free energy change, indicating large hidden dissipation exists. Possible sources of the hidden dissipation were explored by analyzing the Langevin dynamics of the probe, which incorporates the two-state Markov stepper as a kinesin model. We conclude that internal dissipation is dominant.

Kinesin-1 (hereafter called kinesin) is a molecular motor that transports various cargos along microtubules throughout the cell [1,2]. Single molecule kinesin takes 8 nm steps [3] per ATP hydrolysis [4,5] on a microtubule rail and generates ≈7 pN maximum force [6-8]. The two catalytic sites (heads) hydrolyze ATP in a "hand-over-hand" manner that mimics bipedal walking [9-11] by alternating its two heads in coordination with different nucleotide/microtubule binding states [12]. Kinesin shows backward steps occasionally at no load and frequently at high loads [13-15]. Recent experiments indicate that the biased unidirectional motion is achieved by regulating selective binding/unbinding of the head to/from the appropriate binding site [16-20]. Contrary to the molecular mechanism of the motility, the thermodynamic energetics of the motor is poorly understood due to kinesin's stochastic and nonequilibrium behavior.

The energetics of single-molecule motors were historically discussed when their stall forces were measured [21]. Kinesin's stall force of ≈7 pN indicates that maximum work per 8-nm step (≈56 pN·nm) is smaller than the physiological free energy change per ATP hydrolysis (≈85 pN·nm). This is in contrast to rotary motor $F_1$-ATPase, whose stall force explains all input free energy [22]. Kinesin's inefficient work at stall has been regarded as an "open problem" [23]. However, it may not be appropriate to evaluate kinesin efficiency from the stall force for two reasons. First, it is believed that kinesin consumes ATP at backsteps instead of synthesizing ATP [13-15], indicating that the stall condition is not thermodynamically (quasi-)static. Second, the physiological role of kinesin is to carry vesicles against viscous media, meaning that the input energy is dissipated as "heat" rather than "work." Thus, measuring the "dissipation" from the motor is essential when discussing kinesin's nonequilibrium energetics in physiologically relevant conditions.

The Harada-Sasa equality is best suited for this purpose [24,25]:

$$J_x = \gamma \bar{v}^2 + \gamma \int_{-\infty}^{\infty} \left[ \tilde{C}(f) - 2k_B T \tilde{R}'(f) \right] df. \quad (1)$$

Here, $J_x$ is total heat dissipation per unit time from the system through specific degrees of freedom indicated with subscript $x$. $\gamma$ is viscous drag, and $\bar{v} \equiv \langle \dot{x} \rangle$ is mean velocity, where $\langle \cdot \rangle$ denotes the ensemble average. $\tilde{C}(f)$, with frequency $f$, is a Fourier transform of the correlation function of velocity fluctuations, $C(t) \equiv \langle [v(t) - \bar{v}][v(0) - \bar{v}] \rangle$. $\tilde{R}(f)$ is a Fourier transform of the velocity response function, and the prime indicates the real part of the function. $k_B$ is the Boltzmann constant, and $T$ is the absolute temperature. It is known that the fluctuation-response relation (FRR), $\tilde{C}(f) = 2k_B T \tilde{R}'(f)$, is held in equilibrium [26], but the relation is violated in nonequilibrium conditions [27,28]. The integral in Eq. (1) thus indicates heat dissipation that appears only in nonequilibrium systems [29-32]. Since the formula is written with experimentally accessible quantities, Eq. (1) allows us to obtain nonequilibrium heat dissipation, which has been hard to measure directly in small stochastic systems.

Here, we measured the nonequilibrium energy flow of single-molecule walking kinesin via the FRR violation of an attached probe particle using optical tweezers. By analyzing the energy transmission between the probe and kinesin molecule with a mathematical kinesin model, we conclude that the unidirectional motion of kinesin consumes its chemical energy mainly as internal dissipations.

The experimental setup was based on the microscope equipped with optical tweezers described in [33,34] with modifications to incorporate a fast feedback force clamp and epi-fluorescent imaging (Fig. 1a; see also [35] for details). Tail-truncated kinesin constructs [12,35,36] were conjugated to 489 nm probe particles [35,37]. Fluorescent microtubules [35,38] were non-specifically attached to a glass flow cell.

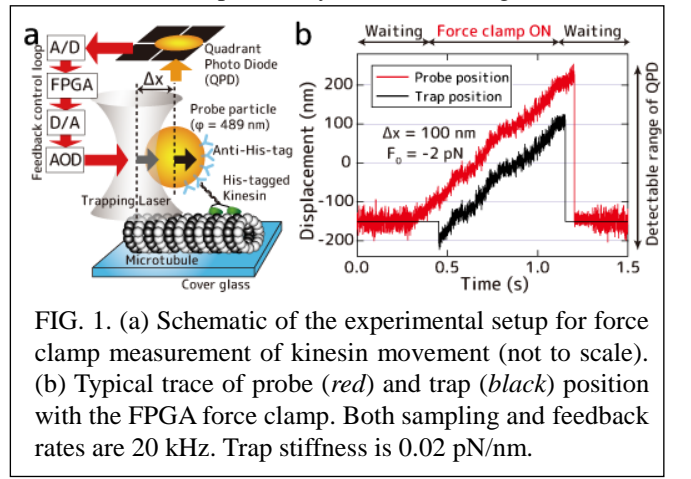

FIG. 1. (a) Schematic of the experimental setup for force clamp measurement of kinesin movement (not to scale). (b) Typical trace of probe (*red*) and trap (*black*) position with the FPGA force clamp. Both sampling and feedback rates are 20 kHz. Trap stiffness is 0.02 pN/nm.

After the cell was washed with casein solution, the probes were trapped on the microtubule at the indicated concentrations of ATP, ADP and potassium phosphate ($P_i$) at 25 ± 1°C. The bright field image of the trapped probe was projected onto quadrant photodiodes (QPD), and the signals were acquired by a field programmable gate array (FPGA)-embedded data acquisition board at a sampling rate of 20 kHz. The feedback-regulated trap positions were calculated from the signals on the FPGA circuit at the same rate, allowing the probe to apply arbitrary force via acousto-optic deflectors (AOD). Displacement calibration was performed by two-dimensional scanning [35,39], where the residual error is <1 nm RMS. Trap stiffness was determined by standard methods [35,40]. Detailed methods and data analysis are described in [35].

Fig. 1b shows single-molecule kinesin movement observed by using force-clamp optical tweezers with FPGA feedback. The apparatus automatically detects the kinesin walking and starts force-clamp mode, which keeps the distance between the probe and the trap center constant. The trap center thus follows the probe motion that displays both thermal fluctuation and kinesin movement until the probe arrives at the end of the detectable range of the QPD. To obtain the response functions, we applied constant forces plus a sinusoidal perturbation of 1/5 times their magnitude [35].

Fig. 2a shows a FRR of the probe movement at high ATP concentration (1 mM ATP, 0.1 mM ADP, 1 mM $P_i$), which simulates physiological conditions where the input free energy change $\Delta\mu = \Delta\mu^0 + k_B T \ln([ATP]/[ADP][P_i])$ is 84.5 ± 2.5 pN·nm [30,41-44]. Constant force was chosen as $F_0 = -2$ pN, which simulates the condition for maximum output power [13]. The response, $2k_B T \tilde{R}'(f)$, and the fluctuation, $\tilde{C}(f)$, took almost the same values at high frequencies, but began to deviate once frequency fell below 20 Hz. This qualitative frequency dependence is consistent with FRR violations reported in prior studies [27,30]. However, the experimental bandwidth of the FRR was limited (Fig. 2a) because of the small observation time period. The kinesin went out of detectable range in a few sub-seconds.

At low ATP concentration (10 μM ATP, 1 μM ADP, 1 mM $P_i$), the kinesin velocity was reduced, while $\Delta\mu$ was kept constant by maintaining the [ATP]/[ADP][$P_i$] ratio. In this condition, the spectra were extended to lower frequencies so that violation of the FRR was observed (Fig. 2b). The nonequilibrium dissipation rate corresponds to the integrated area of the deviation (Fig. 2b, *shaded area*). The dissipation via viscous drag and the output power against the external force are shown in Table 1. The results show that both the nonequilibrium and viscous drag dissipation rates are smaller than the output power by over one order of magnitude.

The relationship between the input $\Delta\mu$ per unit time, the output power and the total heat dissipation rate is provided as

$$\Delta\mu/\tau = -F_0 \bar{v} + J_x + J_{All\ others}, \qquad (2)$$

where $J_{All\ others}$ is the heat dissipation rate via degrees of freedom not observed here, and $\tau$ is the turnover time for ATP hydrolysis by kinesin, which is estimated as $\tau \approx d/\bar{v}$ since the frequency of backsteps at $F_0 = -2$ pN is negligible. The experimental results indicate that the dissipation from the probe's degree of freedom ($J_x$) is dramatically smaller than the power. The sum of these values [the first two terms of the right hand side of Eq. (2)] is ≈20% of the input, $\Delta\mu$ (Table 1), indicating that most (≈80%) of $\Delta\mu$ is dissipated via other *hidden* degrees of freedom ($J_{All\ others}$).

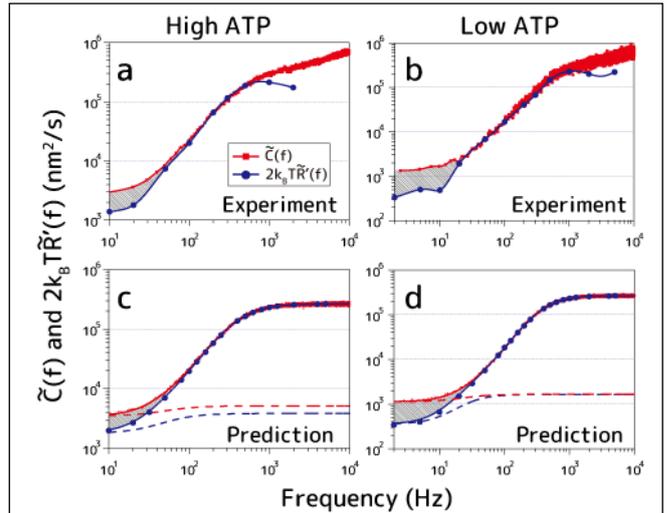

FIG. 2. (a) Typical examples of experimental results of FRR at high ATP (1 mM ATP, 0.1 mM ADP, 1 mM $P_i$) and (b) low ATP (10 μM ATP, 1 μM ADP, 1 mM $P_i$). *Square dots*: velocity fluctuations, *Circles*: response functions. *Lines* are cubic spline interpolations. (c) Model predictions of FRR at high ATP and (d) low ATP. *Dots and circles:* numerical simulations. *Lines* indicate analytical solutions. *Dashed lines* are analytical solutions from the kinesin motor. *Shaded areas* indicate violation of the FRR.

Next we examine the origin of the hidden dissipation using a quantitative theoretical model. Existing mathematical models for kinesin movement fall into two classes. One mimics the kinesin movement using toy models such as thermal ratchets in which the kinesin tumbles on a (switching) one-dimensional potential [45-50]. To date, however, it was reported that when kinesin steps backward kinesin hydrolyzes ATP instead of synthesizing ATP [13-15], indicating that the backward step is not the reverse reaction of the forward one. Therefore, kinesin movement cannot be described by a one-dimensional potential [51]. Instead, kinesin is now believed to take multiple, branched kinetic pathways [17,51]. The other class adopts the Markov chain model to describe the discrete stochastic transitions in the network of kinetic states [23,52-55]. Although single-molecule observations and/or biochemical assays are used to extract the reaction rates between discrete states, most theoretical Markov models require experimentally inaccessible parameters.

Here, to investigate the kinesin movement without parameter tuning, we chose a phenomenological description that only uses experimentally accessible parameters. While abounding (Markov-like) kinetic diagrams have been proposed based on experimental observations

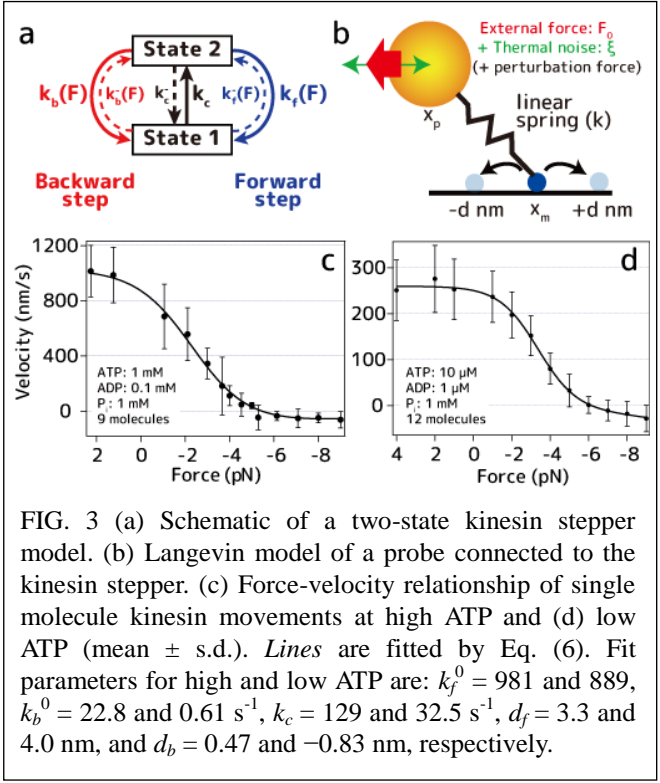

FIG. 3 (a) Schematic of a two-state kinesin stepper model. (b) Langevin model of a probe connected to the kinesin stepper. (c) Force-velocity relationship of single molecule kinesin movements at high ATP and (d) low ATP (mean ± s.d.). *Lines* are fitted by Eq. (6). Fit parameters for high and low ATP are: $k_f^0$ = 981 and 889, $k_b^0$ = 22.8 and 0.61 s$^{-1}$, $k_c$ = 129 and 32.5 s$^{-1}$, $d_f$ = 3.3 and 4.0 nm, and $d_b$ = 0.47 and −0.83 nm, respectively.

[7,8,13,14,17,18,20,51], some intermediate (kinetic) parameters, especially for backsteps, are too rare or transient to determine experimentally [17]. We therefore adopted a simplified kinesin model in which the reaction process was reduced into two-state Markov transitions [14] (Fig. 3a). In this model, the ATP hydrolysis cycle is composed of three transition paths (*solid allows*). One is load independent with rate constant $k_c$ (state 1 to state 2), meaning that the reaction path is not coupled to any mechanical transitions (steps). The second and third paths are load dependent with $k_f$ and $k_b$ (state 2 to state 1) and are coupled to the mechanical transitions for forward and backward steps, respectively. The load dependence of the transition rates are described by Bell's equation [56]:

$$k_f(F) = k_f^0 \exp\left(\frac{d_f F}{k_B T}\right) \quad (3)$$

$$k_b(F) = k_b^0 \exp\left(\frac{d_b F}{k_B T}\right), \quad (4)$$

where $F$ is an external force (load), $k_f^0$ and $k_b^0$ are the rate constants at zero load, and $d_f$ and $d_b$ are the characteristic distances for forward and backward steps, respectively.

To satisfy thermodynamic consistency, *local detailed balance* [57-59] (constraints similar to *microscopic reversibility* [60,61] or *steady-state balance* [62]) conditions are required. For this purpose, three reverse paths are exhibited in the model (*dashed allows* in Fig. 3a). Although small reverse rates for ATP synthesis by kinesin was measured by oxygen isotopic exchange [63], coupling with step movement have not been experimentally identified yet. Furthermore, in our experimental conditions at high Δμ, the reverse rates ($k_c^-$, $k_f^-$ and $k_b^-$) are estimated to be negligibly small and barely affect the model predictions for FRR (*data not shown*). Thus, we use the simplified model that neglected these reverse transitions in the following analysis.

The observable in our experiment is the probe's position, which is pulled by the kinesin motor (Fig. 3b). The motor is modeled as the Markov stepper that transits forward and backward with stepsize $d$, and the probe is connected with the motor via a linear spring of stiffness $k$. The probe's dynamics is described by an overdamped Langevin equation:

$$\gamma \frac{d}{dt} x_p = k(x_m - x_p) + F_0 + N_0 \sin 2\pi f t + \xi, \quad (5)$$

where $x_m$ and $x_p$ are the position of the motor and the probe, respectively, $F_0$ is the external constant force, $N_0$ is the magnitude of the sinusoidal perturbation force for response calculations with frequency $f$, and $\xi$ is the thermal fluctuation force that satisfies $\langle \xi \rangle = 0$ and $\langle \xi(t)\xi(t') \rangle = 2k_B T \gamma \delta(t-t')$.

Five parameters for the kinesin kinetic model, $k_f^0$, $k_b^0$, $k_c$, $d_f$, and $d_b$, were obtained from the experimental results of the force-velocity relationship by fitting the theoretical equation derived from the model (Fig. 3c and d; [35] for deviation):

$$\bar{v} = d \times \frac{(k_f - k_b)k_c}{k_f + k_b + k_c}, \quad (6)$$

where $\bar{v}$ was the mean velocity, and $d$ was the stepsize (8 nm) [3]. The stiffness, $k$ = 0.075 pN/nm, and viscous drag, $\gamma$ = 3.1 × 10$^{-5}$ pN/nm·s, were also obtained experimentally [35].

Analytical solutions for the fluctuation and the response of the model were derived [35] as:

$$\tilde{C}(f) = \frac{k^2 \tilde{C}_m + 2k_B T \gamma \left(k\tilde{R}_m + i2\pi f\right)^2}{\left|k(1+\gamma\tilde{R}_m) + i\gamma 2\pi f\right|^2} \quad (7)$$

$$2k_B T \tilde{R}'(f) = \mathrm{Re}\left[\frac{2k_B T \left(k\tilde{R}_m + i2\pi f\right)}{k(1+\gamma\tilde{R}_m) + i\gamma 2\pi f}\right], \quad (8)$$

with $\tilde{C}_m = d^2 k_c / k_a [(k_f + k_b) - 2(k_f - k_b)^2 k_c / (k_a^2 + 4\pi^2 f^2)]$, $\tilde{R}_m = dk_c / k_a [(\alpha_f - \alpha_b) - (\alpha_f + \alpha_b)(k_f + k_b)/(k_a + i2\pi f)]$, $k_a \equiv k_f + k_b + k_c$, $\alpha_f \equiv k_f d_f / k_B T$, and $\alpha_b \equiv k_b d_b / k_B T$. We confirmed the Harada-Sasa equality (1) holds for the model used here analytically by independently calculating the definition of $J_x \equiv \langle (\gamma v - \xi) \circ v \rangle$ and the right hand side of Eq. (1)[35]. Here, the circle indicates the Stratonovich product [64].

The FRRs obtained from the analytical solutions and numerical simulations are shown in Fig 2c and d (*lines and dots*, respectively). The predicted FRRs similar to the experimental results (Fig. 2a and b) justify our analysis. The obtained dissipation rates, powers and input energy flows are shown in Table 1. In addition, $J_x$ was independently obtained by numerical simulations and the analytical solution. These values are similar to the experimentally obtained values, indicating that the model reproduces the experimental results.

Totally in contrast to the translational motor kinesin studied here, Toyabe *et al.* found that the sum of the work and the dissipations of the rotary motor $F_1$-ATPase were almost the same as the input, $\Delta\mu$, indicating that the internal dissipation of the motor is negligible [30,31]. One candidate reason for the discrepancy is *reversibility*; $F_1$-ATPase acts reversibly as a "power generator" that synthesizes ATP with backward rotation [65,66]. Theoretical models based on the reversibility could successfully explain the little internal dissipation [67,68]. Conversely, kinesin is irreversible and has multiple pathways that include futile ATP hydrolysis such as that for backsteps. The futile ATP hydrolysis *per se* should cause futile dissipation. However, at the experimental condition utilized here, the frequency of the backsteps is only a few percentage points. Although other slippage pathways are also conceivable, they are thought to be rare at conditions similar with this study [69]. This means that futile ATP hydrolysis mainly due to backsteps cannot account for the ≈80% hidden dissipation, at least under physiological conditions.

Another candidate reason is the *softness of the linker* between the probe and the motor. Because of the softness, the non-thermal fluctuation derived from the motor does not transmit to the probe efficiently at high frequencies, and the probe fluctuates merely thermally [27]. In the prior study using $F_1$-ATPase [30,31], the duplex probe is directly connected to the rotary shaft so that the stiffness of the linker is considerably higher than that of kinesin. For kinesin, however, the probe is only loosely connected to the motor via its long stalk; the probe could not completely follow kinesin's rapid steps such that the observed dissipation ($J_x$) might underestimate the actual dissipation from the kinesin movement. We therefore discuss the FRR of the kinesin motor separately from the probe, as explained below.

*Dashed lines* in Fig. 2c and d indicate the fluctuation and the response from the motor obtained using the analytical solution of the two-state Markov model ($\tilde{C}_m$ and $R'_m$; [35] for derivations). At low frequencies, the FRR of the probe nearly agrees with the FRR of the motor, whereas violation of the FRR of the probe seems to attenuate over the cutoff frequency, $f_c = k/2\pi\gamma$. Nevertheless, at low ATP, the FRR violation from the motor approximately agrees with that from the probe, indicating that the probe's FRR almost accurately reports the dissipation from the motor despite the *softness of the linker*. Meanwhile, at high ATP, the FRR violation from the motor was observed even at the highest frequency. (Note that FRR at low ATP also shows small deviations at high frequencies.) The violation of FRR at the high frequency limit is given as $\Delta = dk_c/k_a \left[ k_f(d-2d_f) + k_b(d+2d_b) \right]$ [35,70]. This $\Delta$ originates from the imbalance between the stepsize, $d$, and the characteristic distances, $d_f$ and $d_b$, which is thought to reflect the irreversibility of the system [71], a key element of nonequilibrium dynamics.

When the FRR is violated at high frequency limits, the nonequilibrium dissipation of the motor, which is estimated by integrating the violation towards infinite frequencies, should diverge. This thermodynamic inconsistency appears because the Markov step model assumes that kinesin moves at infinitely large velocity for each step, leading to infinite dissipation. However, the actual kinesin step requires finite time such that the cutoff frequency of the motor movement should exist. Recently, the motion of the kinesin head (≈5 nm) was observed using a small gold particle at the rate of 55 μs (≈18 kHz), where the lag phase during a step could not be resolved [20]. This implies that the cutoff frequency is over tens of kHz and that the FRR violation of the actual motor decays beyond our observation frequencies. We thus made rough estimation of the cutoff frequency and evaluated the dissipation due to whole motor movement, but found we could not completely explain the ≈80% hidden dissipation [21,35].

By dismissing several candidate reasons, the hidden dissipation does not seem to occur through the translational motion of kinesin, but rather is consumed inside the kinesin molecule. The internal dissipation could be explained by introducing additional degrees of freedom in the molecule, whereas our model considered only one observable, i.e. the kinesin position in discrete steps. Experimentally, direct observation of each head revealed the diffusion process, bound/unbound transitions, and off-axis movement of the head [20], suggesting that these microscopic mechanical transitions are required to elucidate the internal dissipations. In addition, the actual kinesin kinetics contains complicated reactions, including inherent *fast* transitions between microscopic structural states [72], whereas our kinesin model used here considered only *slow* transitions between coarse-grained states. Theoretically, these *fast* transitions can also contribute to heat dissipations [73,74]. Although the *fast* transitions neglected in our model mainly consist of intermediate transitions for ATP hydrolysis and they do not directly contribute to the translational motion [17,18,51], dissipations from the *fast* transitions might be required to achieve the biased unidirectional motion. In this case, multi-dimensional mathematical models including reaction coordinates [61,71] could also be suitable for clarifying the total energetics of kinesin.

In summary, we present the first experimental demonstration of violation of the FRR of a single-molecule translational motor, kinesin. By using the Harada-Sasa equality, dissipative heat from the kinesin motor was measured *quantitatively*. The nonequilibrium dissipation via a probe attached to kinesin is dramatically small compared to the output power against external force. The sum of these energy rates is only ≈20% of the input, meaning that most (≈80%) chemical energy is consumed as hidden dissipations, which were not previously recognized. By analyzing the transmission of the motor action to probe fluctuations using a simplified kinesin model, we conclude that the hidden dissipation is "internal dissipation" of the motor. Recently, unobserved reaction pathways, hidden degrees of freedom, and their effects on the energetics of biomolecular machines have been intensively discussed [71,73-75]. By quantifying the internal dissipation of the kinesin molecule, our study will help clarify the unresolved nonequilibrium mechanism.

TABLE I. Summary of experimental results and model predictions for FRR

| Energy flows [pNnm/s] | High ATP Experiment | High ATP Simulation | High ATP Solution | Low ATP Experiment | Low ATP Simulation | Low ATP Solution |
|---|---|---|---|---|---|---|
| $-F_0 \bar{v}$ | 1150 ± 120 | 1110 ± 20 | 1070 | 410 ± 60 | 425 ± 16 | 410 |
| $\gamma \bar{v}^2$ | 10.6 ± 1.9 | 9.58 ± 0.34 | 8.89 | 1.35 ± 0.37 | 1.40 ± 0.11 | 1.29 |
| $2\gamma \int_0^{f_{max}} df \left[ \tilde{C}(f) - 2k_B T \tilde{R}'(\omega) \right]$ | 53.4 ± 41.4[*1] | 27.2[*2] | 50.7 | 2.74 ± 1.52[*1] | 1.83[*2] | 2.14 |
| $J_x$ | 63.9 ± 41.5 | 62.5 ± 6.2[*3] | 59.6 | 4.09 ± 1.56 | 4.53 ± 3.19[*3] | 3.43 |
| $\Delta \mu / \tau$ | 6160 ± 560[*4] | 7000 ± 226 | 6820 | 2190 ± 310[*4] | 2280 ± 105 | 2170 |

[*1] Values were integrated up to $f_{max}$=300 Hz for high ATP (mean ± s.d., $n = 8$) and $f_{max}$=50 Hz for low ATP ($n = 11$).
[*2] Values were integrated up to $f_{max}$=300 Hz for high ATP and $f_{max}$=50 Hz for low ATP.
[*3] Values were directly calculated from the definition of $J_x$ (mean ± s.d., $n = 20$) [64].
[*4] Values were estimated by using the approximation, $\tau \approx d/\bar{v}$.

We acknowledge S.–i. Sasa, K. Kawaguchi and S.-W. Wang for variable discussions, S. Toyabe for helping with data analyzing methods and P. Karagiannis for critically revising the manuscript. This work was supported by JSPS KAKENHI JP25870173, JP15K05248, JP18K03564 (to TA), JP15H01494, JP15H03710, JP25127712 and JP25103011 (to DM).

# Supplemental Material for
# "Nonequilibrium Energetics of Molecular Motor Kinesin"

Takayuki Ariga[1,2], Michio Tomishige[3], Daisuke Mizuno[2]

[1] *Graduate School of Medicine, Yamaguchi University, Yamaguchi 755-8505, Japan*
[2] *Department of Physics, Kyushu University, Fukuoka 819-0395, Japan*
[3] *Department of Applied physics, University of Tokyo, Tokyo 112-8656, Japan*

(Dated: 10 July 2018)

## I. MATERIALS

### A. Kinesin construct

The kinesin construct used in the study is human kinesin-1 'cysteine-light' mutant [1] with a truncated tail region of 490 amino acids to eliminate head-tail interactions [2] (Fig. S1). The construct contained a C-terminal His$_6$ tag (GTHHHHHH), which was used for affinity purifications and also to connect the kinesin to probe particles. Protein expression and purification were carried out as previously described [1].

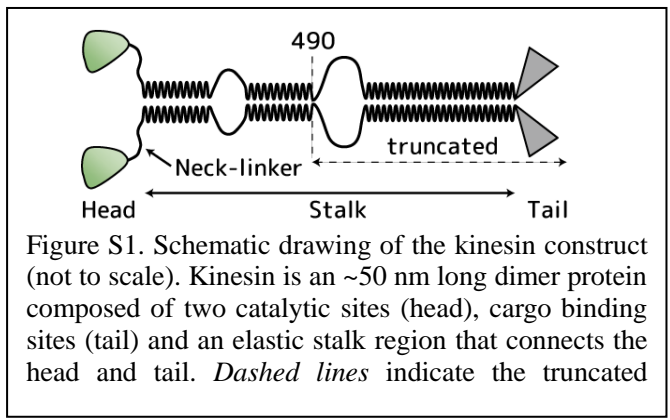

Figure S1. Schematic drawing of the kinesin construct (not to scale). Kinesin is an ~50 nm long dimer protein composed of two catalytic sites (head), cargo binding sites (tail) and an elastic stalk region that connects the head and tail. *Dashed lines* indicate the truncated

### B. Probe particle

Antibody-coated probe particles were prepared by the two-step EDC/sulfo-NHS coupling protocol as described in [3] with slight modifications. 2.69% carboxylate polystyrene beads (diameter, 489 ± 13 nm; Polysciences Inc.) were washed by centrifuging at 14 krpm for 5 min and suspended in MES buffer (50 mM 2-morpholinoethanesulfonic acid (MES)-KOH, pH 5.5, plus 0.1% Tween 20) containing 20 mg/ml 1-ethyl-3-(3-dimethylaminopropyl) carbodiimide (EDC) and 20 mg/ml N-Hydroxysulfosuccinimide (sulfo-NHS). After 15 min incubation at room temperature, the beads were washed two times with MES buffer and resuspended in MES buffer plus 0.14 mg/ml anti 6xHistidine monoclonal antibody (9F2; 010-21861; Wako) and 0.14 mg/ml ATTO532-labeled casein for fluorescence imaging. After 2 h incubation at room temperature, the bead solutions were incubated with 1 mg/ml casein for another 2 h on ice and then mixed with 0.1 M glycine to quench excess reactive sites. The antibody-coated probe particles were washed four times and suspended in BRB80 buffer (80 mM 1,4-piperazinediethanesulfonic acid (PIPES)-KOH, pH 6.8, 2 mM MgCl$_2$, 1 mM EGTA) plus 1 mg/ml casein and 0.1% Tween 20. The probe particles (final concentrations were ~350 pM) were stored at 4°C until use.

Before observation, antibody-coated probe particles and His$_6$-tagged kinesin molecules were mixed for over 3 h on ice at a <1 molecule/particle ratio such that the probability of the movement is less than 30%.

### C. Fluorescent microtubule

Tubulin molecules were purified from pig brains with standard methods using phosphocellulose chromatography [4]. The tubulin molecules were labeled with ATTO532 fluorescent dye as follows. ~6 mg/ml tubulin was polymerized in the presence of 1 mM GTP, 5 mM MgCl$_2$ and 10% dimethyl sulfoxide (DMSO) for 30 min at 37°C. The microtubule solution (~100 μM) was mixed with 1/10 volume of 10 mM ATTO532-NHS ester dye (ATTO-TEC GmbH) and incubated for 10 min at 37°C. 5 mM potassium glutamate was added to terminate the reactions. The solution with labeled microtubules was ultra-centrifuged at 80 krpm, 20 min, 37°C. The pellet was suspended in BRB80 buffer at 0°C and incubated for 10 min at 0°C for depolymerization. After ultra-centrifugation for 5 min at 100 krpm, 2°C, the supernatant was incubated in 1 mM GTP, 5 mM MgCl$_2$ and 10% DMSO for 30 min at 37°C for polymerization and then ultra-centrifuged for 20 min at 80 krpm, 37°C. After repeating the polymerization–depolymerization process again, the pellet was re-suspended in BRB80 buffer at 0°C, incubated for 10 min at 0°C and then ultra-centrifuged at 100 krpm, 5 min, 2°C. The supernatant (ATTO532-labeled tubulin) was frozen in liquid nitrogen and stored at −80°C.

The fluorescently labeled microtubules were prepared as follows. ~1% molar ratio of ATTO532-labeled tubulin was mixed with ~6 mg/ml non-labeled tubulin solution. The mixture was polymerized in the presence of 1 mM GTP, 5 mM MgCl$_2$ and 10% DMSO for 30 min at 37°C. The solution was then mixed with the same amount of BRB80 buffer containing 40 μM paclitaxel, 1 mM GTP and 10% DMSO, incubated 20 min at room temperature and then centrifuged at 14 krpm, 5 min, 25°C. The pellet was rinsed and suspended with BRB80 buffer containing 20 μM paclitaxel, 1 mM GTP and 10% DMSO. The obtained ATTO532-labeled microtubules can be stored in dark at room temperature for several weeks.

## II. METHODS

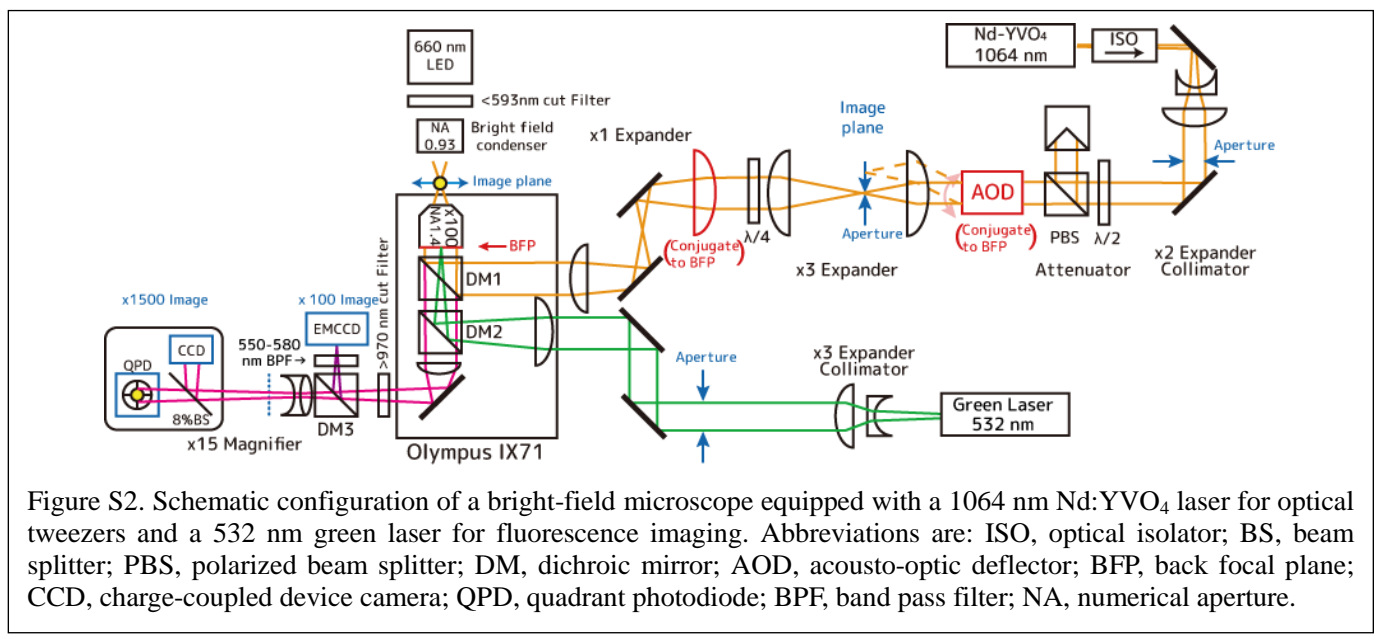

Figure S2. Schematic configuration of a bright-field microscope equipped with a 1064 nm Nd:YVO$_4$ laser for optical tweezers and a 532 nm green laser for fluorescence imaging. Abbreviations are: ISO, optical isolator; BS, beam splitter; PBS, polarized beam splitter; DM, dichroic mirror; AOD, acousto-optic deflector; BFP, back focal plane; CCD, charge-coupled device camera; QPD, quadrant photodiode; BPF, band pass filter; NA, numerical aperture.

### A. Microscope

The optical tweezers apparatus (Fig. S2) was set on a vibration-free table (HELTZ). An infrared Nd:YVO$_4$ laser (1064 nm, BL-106-TU-E, Spectra-Physics) for optical trapping and a green laser (532 nm, DJ532-10, Thor labs) for fluorescence imaging were brought into an inverted bright field microscope (IX71, Olympus). The trap laser beam was expanded and collimated to overfill the back aperture of an oil immersion objective lens (UplanSApo, 100×, NA 1.4, Olympus) by going through an optical isolator (ISO; IO-3-1064-VHP), a Galilean beam expander, a power attenuator, two Keplerian beam expanders and a dichroic mirror (DM1) [5]. ATTO532 labeled microtubules and probe particles were excited by epi-fluorescent illumination with the green laser, and the fluorescent images were monitored with an electron multiplying charge coupled device camera (EMCCD; MC681SPd-R0B0, Texas Instruments). The bright field image of the probe particle, which was illuminated with a high power LED light (M660L3, Thor labs), was ≈1500 times magnified and projected onto a quadrant photodiode (QPD; S4349, Hamamatsu Photonics). The voltage outputs of the QPD were amplified by a differential amplifier (OP711A, Sentech) and recorded onto a field programmable gate array (FPGA) embedded data acquisition board (NI PCIe-7842R, National Instruments) at a sampling rate of 20 kHz. The acquired signals were converted to the *x-y* displacements by a fifth-order polynomial calculation with 72 parameters [6] on the FPGA circuit that outputs the feedback signals to control the laser position at the same rate of 20 kHz. The output signals were applied to two analog RF drivers (DE-272JM, IntraAction) for the 2-axis acousto-optic deflectors (AOD; DTD-274HD6M, IntraAction), which control the angle of the laser beam by changing the acoustic frequency. The AOD was located at the conjugate to the back focal plane of the objective lens. This optical layout allows control of the *x-y* position of the laser focus by the beam angle at the AOD.

### B. Observation flow cell

A flow cell (~5 μl volume) was constructed between two plasma-cleaned (or KOH-cleaned) cover glasses (24 mm × 36 mm and 18 mm × 18 mm; Matsunami Glass Ind.) by placing between them two spacers of ~50 μm thickness that coated with silicon grease. ATTO532 labeled microtubules diluted with BRB12 buffer (12 mM PIPES-KOH, pH 6.8, 2 mM MgCl$_2$, 1 mM EGTA) plus 20 μM paclitaxel were infused into the flow cell. By incubating for 3 min, microtubules were fixed non-specifically on the glass surface. The cell was washed with BRB12 buffer including 1 mg/ml casein and 20 μM paclitaxel to remove unbound microtubules. The cell was then washed with ~1 pM kinesin-coated probe solution, which contained 12 mM PIPES-KOH (pH 6.8), 2 mM MgCl$_2$, 1 mM EGTA, 29 mM potassium acetate, 50 U/ml glucose oxidase, 50 U/ml catalase, 4.5 mg/ml glucose, 0.5% 2-mercaptoethanol, 0.4 mg/ml casein, 20 μM paclitaxel and the indicated concentrations of ATP, ADP and potassium phosphate (P$_i$). The cell was sealed with silicon grease to avoid evaporation. These procedures were performed at room temperature (~25°C).

### C. Single molecule manipulation

Probe particles floating in the flow cell were monitored by epi-fluorescence imaging and trapped by the optical tweezers. The trapped probe was put close to a microtubule that was laid fixed on the grass surface in a direction according to the *x*-axis of the QPD. After checking the interaction between the kinesin and the microtubule by observing the probe movement and its direction, calibration of the position and the trap stiffness were performed for each probe. Then, the force-clamp mode was turned on, where the trap position and the trap stiffness were set to 100 nm behind the probe center and 0.02 pN/nm (for $F_0 = -2$ pN), respectively. The probe was put on standby at the starting-end of the calibrated range of QPD (typically ±210 nm).

The movement of a probe pulled by kinesin walking on the microtubule was automatically detected, and then constant force was applied via the force clamp (Fig. 1b in the main text). When the probe reached the other end of the calibrated range or detached from the microtubule, the feedback loop was automatically stopped, and the trap stiffness increased to force the probe to return to the waiting position. The probe was then set back to the adequate trap stiffness and restarted the force-clamp measurement. This cycle was repeated several times to obtain the velocity fluctuation, $\tilde{C}(f)$, for each probe.

To obtain the response function, $2k_BT\tilde{R}'(f)$, a sinusoidal displacement with amplitude 20 nm was added to the displacement between the trap and the probe position by changing the frequency such that the sinusoidal perturbation force, $N_0$, was 1/5 the magnitude of the constant force, $F_0$ (i.e. $N_0 = 0.4$ pN for $F_0 = -2$ pN). To obtain the force-velocity relationship, the trap stiffness was changed from 0.01 to 0.09 pN/nm, while the distance between the probe and the trap center was fixed to ± 100 nm. Although the probe movement occasionally showed over 16 nm jumps, we excluded the trajectory from further analysis because such a large jump seemed more consistent with the kinesin detachment and re-attachment process than steady-state walking steps. In addition, probes that obviously showed multi-molecular behavior (e.g. multistep detachment or >8 pN stall forces) were also omitted. All manipulation and data acquisition were done with custom-written LabVIEW programs. The probe and trap position data were recorded on a PC. All observations were performed at 25 ± 1°C.

### D. Trap stiffness and displacement calibration

Calibration for the trap stiffness was performed according to the method by Gittes et al. [7]. For the measurement, trap stiffness of the optical tweezers ($k_{trap}$) is determined from the variance of the trapped probe beads $\langle x^2 \rangle$ by using the equipartition law,

$$\frac{1}{2}k_{trap}\langle x^2 \rangle = \frac{1}{2}k_BT. \quad (S.9)$$

Since the above calibration method tends to be affected by systematic noise, we evaluated the accurate trap stiffness for further analysis as follows. The power spectrum density (PSD) of the probe position, $S(f)$, was fitted by a Lorenzian function:

$$S(f) = \frac{S_0}{1 + f^2/f_c^2}, \quad (S.10)$$

where the horizontal line, $S_0$, and the corner frequency, $f_c$, were determined. The accurate value of the trap stiffness was then calculated as

$$k_{trap} = \frac{2k_BT}{\pi S_0 f_c}. \quad (S.11)$$

Calibration for the displacement was performed by two-dimensional (2D) scanning over a 420 × 420 nm squared area with fifth-order polynomial fitting [6], where the residual error of the fit is <1 nm RMS.

### E. Data analysis to calculate FRR

Experimental data were analyzed following the procedure described in Toyabe et al. [8] with slight modifications to adapt to the kinesin motor. To obtain the response, $2k_BT\tilde{R}'(f)$, we applied sinusoidal perturbation force to the probe as

$$N(t) = N_0 \sin(2\pi f_0 t), \quad (S.12)$$

where $f_0$ is a constant frequency. The measured velocity of the probe was averaged synchronously with respect to the perturbation and fitted with the sinusoidal function $v(t) = \bar{v} + A_0 \sin(2\pi f_0 t + \phi)$, where $\bar{v}$ is the steady state velocity, $A_0$ is the amplitude of the velocity perturbation and $\phi$ is the phase shift. From the parameters, we obtained the real part of the response function as

$$\tilde{R}' = \frac{A_0}{N_0}\cos(\phi). \quad (S.13)$$

The instantaneous velocity includes noise, such as thermal fluctuations, if it is obtained by simply differentiating the displacement data. Synchronous averaging was not enough to remove noise for small $f_0$, since the total number of periods averaged cannot be large (<100 periods for $f_0 = 2$ Hz). To further reduce noise, prior to the velocity calculation, the displacement data was processed by a median filter that smooths with 1/100 the sampling number of each period to obtain the response at $f_0 \leq 50$ Hz.

The velocity fluctuation, $\tilde{C}(f)$, was calculated as a PSD from the probe velocity measured under constant external force. The velocity was obtained without any smoothing, and the PSD was calculated by a fast Fourier transform using a Hanning window.

The nonequilibrium dissipation rate was obtained by calculating the integral in the Harada-Sasa equality [9]:

$$J_x = \gamma\bar{v}^2 + \gamma\int_{-\infty}^{\infty}\left[\tilde{C}(f) - 2k_BT\tilde{R}'(f)\right]df. \quad (S.14)$$

In our experiments, both $\tilde{C}(f)$ and $\tilde{R}'(f)$ decay several orders of magnitude at low frequencies compared with high frequencies. Therefore, small deviations of $\tilde{C}(f)$ and $2k_BT\tilde{R}'(f)$ at high frequencies lead to significant errors in the estimate of the nonequilibrium dissipation, though the FRR should hold [8]. Data at high frequencies are also affected by systematic errors, such as aliasing, feedback delay on the FPGA circuit, electric and shot noise, etc. Therefore, we introduced a cutoff, $f_{max}$, in the integration in Eq. (S. 6) as

$$2\gamma\int_0^{f_{max}}\left[\tilde{C}(f) - 2k_BT\tilde{R}'(f)\right]df. \quad (S.15)$$

$f_{max}$ was determined such that the standard deviation of the dissipation does not exceed the mean value. That is, 300 Hz for samples at high ATP and 50 Hz at low ATP.

All data analysis and model simulations were performed by custom-written Igor Pro 6.3 64-bit procedures.

### F. Parameter estimation for the model

To simulate the kinesin movement, it was necessary to determine the spring constant (stiffness), $k$, that connects the probe to the motor head (*e.g.* kinesin stalk) and viscous drag coefficient, $\gamma$, of the probe particle, both of which were measured as follows.

The spring constant, $k$, was obtained by measuring the thermal fluctuation of the probe particle. Via the coated kinesin with non-hydrolyzable nucleotide, AMP-PNP, the probe particle was stably bound to the microtubule. Then we obtain $k$ from the PSD of the probe displacements (Fig. S3a) by fitting Eq. (S.10) and by using

$$k = \frac{2k_B T}{\pi S_0 f_c}. \quad (S.16)$$

Spring constants measured under various constant loads are shown in Fig. S3b. The spring constants depend on the externally applied force, indicating nonlinear behavior. We therefore utilized the value $k(F_0)$, which corresponds to the force, $F_0$, externally applied to the probe. For example, $k(F_0 = -2$ pN$) = 0.075 \pm 0.012$ pN/nm (mean ± s.d., $n = 6$). As we see below, force applied to the stalk is further perturbed by the abrupt step of kinesin and the thermally fluctuating forces. We neglected the effect of these small perturbations on $k$ to evaluate the FRR.

The viscous drag coefficient, $\gamma$, was obtained from the same PSD [7] as

$$\gamma = \frac{k_B T}{\pi^2 S_0 f_c^2}. \quad (S.17)$$

The obtained $\gamma$ value is $(3.09 \pm 0.80) \times 10^{-5}$ pN/nm·s (mean ± s.d., $n = 6$) at $F_0 = -2$ pN.

### III. ANALITYCAL SOLUTIONS

The FRRs of kinesin movement and the probe attached to it were investigated. We first evaluate the linear response to small perturbations and then derive velocity fluctuations, both under constant external force. At each step, we first analyze the two-state Markov model for kinesin and incorporate its solution to the Langevin dynamics of the probe (Fig. 3a and b in the main text).

### A. Response function of the kinesin motor

Kinesin's transition between state 1 and state 2 is given by

$$\text{State 1} \xrightarrow{k_c} \text{State 2} \xrightarrow{k_f + k_b} \text{State 1}. \quad (S.18)$$

The probability to dwell in each state ($P_1$ and $P_2$) obeys the following master equation:

$$\frac{d}{dt} P_2 = k_c P_1 - (k_f + k_b) P_2, \quad (S.19)$$

where $k_c$ is load independent, and $k_f$ and $k_b$ are load dependent rate constants [10]. The load dependency is given as

$$\begin{aligned} k_f &= k_f^0 \exp\left(\frac{d_f F}{k_B T}\right) \\ k_b &= k_b^0 \exp\left(\frac{d_b F}{k_B T}\right), \end{aligned} \quad (S.20)$$

where $F$ is an external force, $k_f^0$ and $k_b^0$ are the rate constants at zero load, and $d_f$ and $d_b$ are the characteristic distances for forward and backward steps, respectively. For steady state condition ($dP_2/dt = 0$), using the conservation of the probability $P_1 + P_2 = 1$, we obtain the mean probability at each state as:

$$\begin{aligned} \bar{P}_1 &= \frac{k_f + k_b}{k_f + k_b + k_c} \\ \bar{P}_2 &= \frac{k_c}{k_f + k_b + k_c}. \end{aligned} \quad (S.21)$$

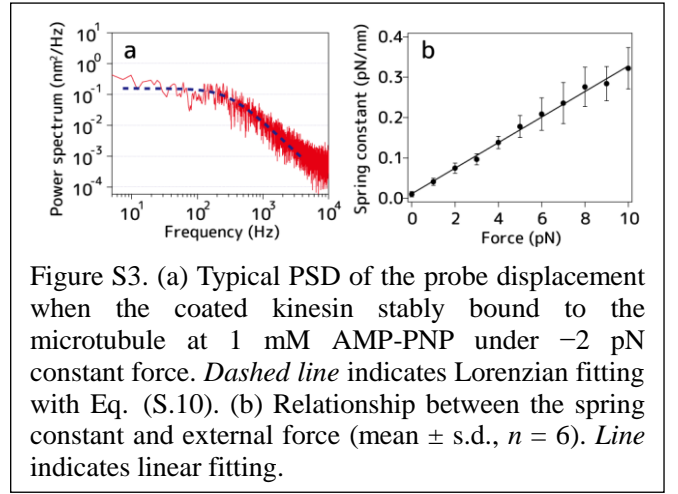

Figure S3. (a) Typical PSD of the probe displacement when the coated kinesin stably bound to the microtubule at 1 mM AMP-PNP under −2 pN constant force. *Dashed line* indicates Lorenzian fitting with Eq. (S.10). (b) Relationship between the spring constant and external force (mean ± s.d., $n = 6$). *Line* indicates linear fitting.

Hereafter, the bar above a parameter indicates a steady-state quantity. The steady-state velocity of kinesin, $\bar{v}_m$, is related to the transition probability, $P_2$, as

$$\bar{v}_m = d \times (k_f - k_b) \bar{P}_2, \quad (S.22)$$

where $d$ is the stepsize. By substituting Eq. (S.21) into Eq. (S.22), we obtain the steady-state velocity of the kinesin movement (Eq. (5) in the main text) that was fit to the force-velocity relationship in order to obtain the kinetic parameters $k_f^0$, $k_b^0$, $k_c$, $d_f$ and $d_b$.

Let the total force, $F_m$, applied to the kinesin motor be divided into mean force, $\bar{F}_m$, and perturbation force, $\delta F_m$, as

$$F_m = \bar{F}_m + \delta F_m. \quad (S.23)$$

When $\delta F_m$ is added to the external force, the load-dependent rate constants of the kinesin motor are given by the first-order approximation as

$$\begin{aligned} k_f &= \bar{k}_f + \delta k_f = \bar{k}_f + \alpha_f \delta F_m \\ k_b &= \bar{k}_b + \delta k_b = \bar{k}_b + \alpha_b \delta F_m, \end{aligned} \quad (S.24)$$

where $\alpha_f$ and $\alpha_b$ are defined by

$$\alpha_f \equiv k_f^0 \exp\left(\frac{\bar{F}_m d_f}{k_B T}\right) \times \frac{d_f}{k_B T}$$
$$\alpha_b \equiv k_b^0 \exp\left(\frac{\bar{F}_m d_b}{k_B T}\right) \times \frac{d_b}{k_B T}. \tag{S.25}$$

When $\delta F_m$ is applied to the kinesin motor, the probability in state 2, $P_2$, is also deviated from the steady state probability, $\bar{P}_2$. We expand the master equation (S.19) to the linear order, $\delta P_2$ and $\delta F_m$, by using Eq. (S.24) and $P_2 = 1 - P_1 = \bar{P}_2 + \delta P_2$ as

$$\frac{d}{dt}(\bar{P}_2 + \delta P_2) = k_c(1 - \bar{P}_2 - \delta P_2) - (\bar{k}_f + \alpha_f \delta F_m + \bar{k}_b + \alpha_b \delta F_m)(\bar{P}_2 + \delta P_2). \tag{S.26}$$

By using the relation $d/dt\, \bar{P}_2 = k_c \bar{P}_1 + (\bar{k}_f + \bar{k}_b)\bar{P}_2 = 0$, we have

$$\frac{d}{dt}\delta P_2 = -k_a \delta P_2 - (\alpha_f + \alpha_b)\bar{P}_2 \delta F_m, \tag{S.27}$$

where the term $O(\delta F_m^2)$ is omitted for linear approximations, and $k_a \equiv \bar{k}_f + \bar{k}_b + k_c$ is used. Similarly, by using Eq. (S.24) and the steady-state velocity, $\bar{v}_m$, expressed by Eq. (S.22), we obtain the linear response of the velocity to the perturbation force, $\delta F_m$, as

$$\delta v_m = d \times \left[(\alpha_f - \alpha_b)\bar{P}_2 \delta F_m + (k_f - k_b)\delta P_2\right]. \tag{S.28}$$

Applying the sinusoidal perturbation $\delta F_m \propto \exp(i\omega t)$ with the angular frequency $\omega = 2\pi f$, we obtain $\delta P_2$ from Eq. (S.27):

$$\delta P_2 = -\frac{(\alpha_f + \alpha_b)\bar{P}_2}{i\omega + k_a}\delta F_m. \tag{S.29}$$

Then, substituting Eq. (S.29) into Eq. (S.28), we have

$$\delta v_m = \left(A + \frac{B}{i\omega + k_a}\right)\delta F_m \equiv \tilde{R}_m \delta F_m, \tag{S.30}$$

where $A \equiv d \times (\alpha_f - \alpha_b)\bar{P}_2$ and $B \equiv -d \times (\alpha_f + \alpha_b)(k_f - k_b)\bar{P}_2$. $\tilde{R}_m$, seen in Eq. (S.22), is the (complex) velocity response function of the kinesin motor.

### B. Response function of the probe

The equation of motion for the probe is given by

$$\gamma \frac{d}{dt} x_p = k(x_m - x_p) + F_p + \xi \tag{S.31}$$

where $x_m$ and $x_p$ are the position of the kinesin motor and the probe, respectively, $F_p$ is the external force on the probe, and $\xi$ is thermal fluctuation. Hereafter, the subscripts $p$ and $m$ indicate the probe and kinesin motor, respectively. When the perturbation force, $\delta F_p$, is applied to the probe, the change in the linear order is given by

$$\gamma \frac{d}{dt}\delta x_p = k(\delta x_m - \delta x_p) + \delta F_p. \tag{S.32}$$

Here, $\delta x_m$ and $\delta x_p$ are defined as the deviations from the steady-state positions $\bar{x}_m$ and $\bar{x}_p$, respectively. $\xi$ is thus negligible in (S.32). By applying sinusoidal perturbation $\delta F_p \propto \exp(i\omega t)$, we obtain

$$\delta x_p = \frac{k \delta x_m + \delta F_p}{i\omega \gamma + k}. \tag{S.33}$$

By substituting Eq. (S.30), we have

$$\delta v_p = i\omega \delta x_p = \frac{k \tilde{R}_m \delta F_m + i\omega \delta F_p}{i\omega \gamma + k}. \tag{S.34}$$

The relationship between $\delta F_m$ and $\delta F_p$ can be obtained from the relation $\delta F_m = k(\delta x_p - \delta x_m)$ and Eq. (S.32) as

$$\delta F_m = \delta F_p - \gamma \delta v_p \tag{S.35}$$

By substituting Eq. (S.35) into Eq. (S.34), we finally obtain

$$\delta v_p = \frac{k \tilde{R}_m + i\omega}{i\omega \gamma + k(1 + \gamma \tilde{R}_m)}\delta F_p \equiv \tilde{R}_p \delta F_p. \tag{S.36}$$

$\tilde{R}_p$ is the (complex) linear response function for the probe's velocity.

In order to estimate the Harada-Sasa equality (S.14), we utilize the real part of $\tilde{R}_p$ times $2k_B T$, which is calculated as

$$2k_B T \tilde{R}_p' = 2k_B T \frac{k^2 R_m'(1 + \gamma R_m') + \omega^2 \gamma (1 + k R_m'')^2}{k^2(1 + \gamma R_m')^2 + \omega^2 \gamma^2 (1 + k R_m'')^2}, \tag{S.37}$$

where the complex response function of the kinesin motor is expanded as $\tilde{R}_m \equiv R_m' + i\omega R_m''$ with

$$R_m' = A + \frac{k_a B}{\omega^2 + k_a^2}$$
$$R_m'' = \frac{-B}{\omega^2 + k_a^2}. \tag{S.38}$$

### C. Velocity fluctuation of the kinesin motor

Next we evaluate the velocity fluctuation of the kinesin motor. The velocity of the kinesin movement with stepsize $d$ is expressed as the time series of the $\delta$ functions as

$$v_m(t) = \sum_i d \cdot \sigma(i) \delta(t - t_i), \tag{S.39}$$

where $t_i$ is the time for the step to occur, and $\sigma(i) = 1$ for forward steps and $\sigma(i) = -1$ for backsteps. In Fourier space, we have

$$\tilde{v}_m(\omega) = \int d \sum_i \sigma(i) \delta(t - t_i) \exp(-i\omega t) dt$$
$$= d \sum_i \sigma(i) \exp(-i\omega t_i) \tag{S.40}$$

The velocity correlation is divided into two terms as

$$\langle \tilde{v}_m^*(\omega) \tilde{v}_m(\omega) \rangle = \left\langle \sum_{i \neq j} d^2 \sigma(i) \sigma(j) \exp\left[-i\omega(t_i - t_j)\right] \right\rangle$$
$$+ \left\langle \sum_{i=j} d^2 \exp\left[-i\omega(t_i - t_j)\right] \right\rangle \tag{S.41}$$
$$\equiv \tilde{C}_1(\omega) + C_0.$$

Because there are forward and backward steps, $\tilde{C}_1(\omega)$ is composed of 4 combinations of step pairs: forward/forward, backward/backward, forward/backward and backward/forward. The probability for the occurrence of each case is $k_f^2/(k_f+k_b)^2$, $k_b^2/(k_f+k_b)^2$, $k_f k_b/(k_f+k_b)^2$, $k_b k_f/(k_f+k_b)^2$, respectively. Note that it is assumed that the back and forward steps are randomly chosen with the ratios $k_b$ and $k_f$, respectively, when a step occurs. We therefore obtain

$$\tilde{C}_1(\omega) = d^2 \frac{k_f^2 + k_b^2 - 2k_f k_b}{(k_f + k_b)^2} \tilde{f}(\omega), \quad \text{(S.42)}$$

where $\tilde{f}(\omega) \equiv \left\langle \sum_{i \neq j} \exp[-i\omega(t_i - t_j)] \right\rangle$ is a Fourier transform of the pair correlation function between distinct steps, i.e. $f(t) = \left\langle \sum_{i \neq j} \delta(t - (t_i - t_j)) \right\rangle$. We derive $f(t)$ as follows.

Consider a kinesin motor that is in state 1 at time $t_1$. From the master Eq. (S.19), the conditional probability for the same motor being in state 2 at time $t_2$ ($t_2 > t_1$) is

$$P_2(t_2; t_1) = \frac{k_c}{k_a}\{1 - \exp[-k_a(t_2 - t_1)]\}. \quad \text{(S.43)}$$

With the steady state assumption, the probability for the motor in state 2 at any time (and therefore at time $t_1$) is given as

$$P_2(t_1) = \overline{P}_2 = \frac{k_c}{k_a}. \quad \text{(S.44)}$$

Thus, the frequency that a motor steps both at $t_1$ and $t_2$ is

$$f_2(t_1, t_2) = (k_f + k_b) P_2(t_2; t_1)(k_f + k_b) P_2(t_1)$$
$$= \frac{k_c^2 (k_f + k_b)^2}{k_a^2}\{1 - \exp[-k_a(t_2 - t_1)]\}. \quad \text{(S.45)}$$

We then obtain the probability for the *distinct* pair of steps occurring at time $t$ as

$$f(t) = \frac{k_c^2 (k_f + k_b)^2}{k_a^2}\{1 - \exp[-k_a |t|]\}. \quad \text{(S.46)}$$

Note that the first term of Eq. (S.46) corresponds to the steady-state pair correlation function between steps:

$$\overline{f}(t) \equiv \left[(k_f + k_b)\overline{P}_2\right]^2 = \frac{k_c^2 (k_f + k_b)^2}{k_a^2}. \quad \text{(S.47)}$$

Thus, the Fourier transform of (S.46) can be written as

$$\tilde{f}(\omega) = \tilde{\overline{f}}(\omega) - \frac{k_c^2 (k_f + k_b)^2}{k_a^2} \frac{2k_a}{k_a^2 + \omega^2}. \quad \text{(S.48)}$$

Substituting Eq. (S.48) into Eq. (S.42), we obtain the velocity correlation of the distinct pair of steps as

$$\tilde{C}_1(\omega) = \tilde{\overline{v}}_m^2 - \frac{2d^2 (k_f - k_b)^2 k_c^2}{k_a(k_a^2 + \omega^2)}, \quad \text{(S.49)}$$

where we used the steady-state velocity correlation:

$$\overline{v}_m^2 = d^2 \frac{(k_f - k_b)^2 k_c^2}{k_a^2}$$
$$= d^2 \frac{k_f^2 + k_b^2 - 2k_f k_b}{(k_f + k_b)^2} \overline{f}(t). \quad \text{(S.50)}$$

On the other hand, $C_0$ in Eq. (S.41) is proportional to the frequency of the steps per unit time, $(k_f + k_b)\overline{P}_2$, and we obtain

$$C_0 = \frac{d^2 k_c (k_f + k_b)}{k_a} \quad \text{(S.51)}$$

The correlation function of velocity fluctuations is defined as $C(t) \equiv \langle [v(t) - \overline{v}][v(0) - \overline{v}] \rangle$. Thus, the Fourier transform of the velocity fluctuation of the kinesin motor is obtained by using Eq. (S.49) and Eq. (S.51) as

$$\tilde{C}_m(\omega) \equiv \left\langle [\tilde{v}_m^*(\omega) - \tilde{\overline{v}}_m^*][\tilde{v}_m(\omega) - \tilde{\overline{v}}_m] \right\rangle$$
$$= \langle \tilde{v}_m^*(\omega)\tilde{v}_m(\omega) \rangle - \tilde{\overline{v}}_m^2$$
$$= \tilde{C}_1(\omega) + C_0 - \tilde{\overline{v}}_m^2 \quad \text{(S.52)}$$
$$= d^2 \left[ \frac{k_c(k_f + k_b)}{k_a} - \frac{2(k_f - k_b)^2 k_c^2}{k_a(k_a^2 + \omega^2)} \right].$$

It should be noted that the derivation is based on *static* kinetic parameters, meaning that only constant external force application is assumed, that is, fluctuation of the external force applied to the motor is not considered.

## D. Velocity fluctuation of the probe

Next, we derive the velocity fluctuation of the probe attached to a kinesin motor walking along a microtubule. The Langevin Eq. (S.31) is transformed as

$$\gamma \tilde{v}_p = k(\tilde{x}_m - \tilde{x}_p) + \tilde{F}_p + \tilde{\xi}. \quad \text{(S.53)}$$

The velocity fluctuation was measured at the condition of a constant force application to the probe, $F_p = F_0$. However, the total force applied to the motor, $F_m$, depends on the stochastic variables $\xi$ and $v_p$ as follows:

$$F_m = -k(x_m - x_p) = F_p + \xi - \gamma v_p \equiv F_0 + \delta F_m \quad \text{(S.54)}$$

Because the kinesin position, $x_m$, is stochastically determined depending on $F_m$, so too is the velocity $v_m$. We thus divide $v_m$ into two terms: $v_m(t; F_0)$, which is independent of the force fluctuation, and $\delta v_m(t; \delta F_m)$, which is the deviation due to the fluctuation of $F_m$.

$$\tilde{v}_m(\omega; F_m) = \tilde{v}_m(\omega; F_0) + \delta \tilde{v}_m(\omega; \delta F_m)$$
$$= \tilde{v}_m(\omega; F_0) + \tilde{R}_m \delta \tilde{F}_m + O(\delta \tilde{F}_m^2) \quad \text{(S.55)}$$
$$\approx \tilde{v}_m(\omega; F_0) + \tilde{R}_m(\tilde{\xi} - \gamma \tilde{v}_p).$$

In the third line, we have used the linear approximation. Using $\tilde{v} = i\omega\tilde{x}$, we have

$$\tilde{x}_m(\omega; F_m) \approx \tilde{x}_m(\omega; F_0) + \tilde{R}_m\left(\frac{\tilde{\xi}}{i\omega} - \gamma\tilde{x}_p\right). \quad (S.56)$$

Substituting Eq. (S.56) into Eq. (S.53), we obtain the Fourier transformed velocity profile of the probe as

$$\tilde{v}_p(\omega) = \frac{k\tilde{v}_m(\omega; F_0) + (k\tilde{R}_m + i\omega)\tilde{\xi} + i\omega\tilde{F}_p}{k(1+\gamma\tilde{R}_m) + i\omega\gamma} \quad (S.57)$$

Similarly the steady-state velocity of the probe can be obtained using ensemble averages of Eq. (S.53) and Eq. (S.56) as

$$\bar{\tilde{v}}_p = \frac{k\bar{\tilde{v}}_m + i\omega\tilde{F}_p}{k(1+\gamma\tilde{R}_m) + i\omega\gamma}. \quad (S.58)$$

The velocity fluctuations of the probe is

$$\tilde{C}_p(\omega) \equiv \left\langle \left[\tilde{v}_p^*(\omega) - \bar{\tilde{v}}_p^*\right]\left[\tilde{v}_p(\omega) - \bar{\tilde{v}}_p\right] \right\rangle$$

$$= \frac{k^2\tilde{C}_m + (k\tilde{R}_m + i\omega)^2 \tilde{\xi}^2}{|k(1+\gamma\tilde{R}_m) + i\omega\gamma|^2} \quad (S.59)$$

$$= \frac{k^2\tilde{C}_m + 2k_BT\gamma\left[\omega^2(1+kR_m'')^2 + k^2R_m'^2\right]}{k^2(1+\gamma R_m')^2 + \omega^2\gamma^2(1+kR_m'')^2}.$$

Here, the fluctuation-dissipation theorem of the second kind, $\tilde{\xi}^2 = 2k_BT\gamma$, is utilized. Note that $\tilde{C}_m(\omega)$ in this equation denotes the velocity fluctuation of the motor under constant force application, $\left\langle \left[\tilde{v}_m^*(\omega; F_0) - \bar{\tilde{v}}_m^*\right]\left[\tilde{v}_m(\omega; F_0) - \bar{\tilde{v}}_m\right]\right\rangle$, which was already investigated in the previous section [Eq. (S.52)]. Precision of this analytical solution depends on the linear approximation used in deriving Eq. (S.55). The approximation error is, however, negligible in our experimental conditions $\gamma \ll 1$ and $k \ll 1$.

### E. Derivation for Harada-Sasa Equality

Here, we evaluate the lefthand side of the Harada-Sasa equality (S.14) for our simplified kinesin model. The definition of the total dissipation rate, $J_x$, is given by

$$J_x \equiv \left\langle (\gamma v_p - \xi) \circ v_p \right\rangle = \gamma\langle v_p^2 \rangle - \langle \xi \circ v_p \rangle, \quad (S.60)$$

where the circle indicates the Stratonovich product [11]. We have already obtained the Fourier transformed velocity, $\tilde{v}_p$, Eq. (S.57). Thus, we utilize the general relationship between the cross correlation function $C_{xy}(\tau) \equiv \langle x(t)y(t-\tau)\rangle$ and cross spectrum $S_{xy} \equiv \langle \tilde{x}^*(\omega)\tilde{y}(\omega)\rangle$. When $\tau = 0$, the relationship

$$C_{xy}(0) = \int_{-\infty}^{\infty} S_{xy}(\omega)\frac{d\omega}{2\pi} \quad (S.61)$$

is always satisfied. By using this relationship and the Fourier transformed velocity obtained by Eq. (S.57), the first term of Eq. (S.60) is given as

$$\gamma\langle v_p^2 \rangle = \gamma\int_{-\infty}^{\infty} \langle \tilde{v}_p^*(\omega)\tilde{v}_p(\omega)\rangle \frac{d\omega}{2\pi}$$

$$= \gamma\int_{-\infty}^{\infty}\left\langle \bar{\tilde{v}}_p^2 + \left[\tilde{v}_p^*(\omega) - \bar{\tilde{v}}_p^*\right]\left[\tilde{v}_p(\omega) - \bar{\tilde{v}}_p\right]\right\rangle\frac{d\omega}{2\pi} \quad (S.62)$$

$$= \gamma\bar{v}_p^2 + \gamma\int_{-\infty}^{\infty}\tilde{C}_p\frac{d\omega}{2\pi},$$

where we utilized the definition of $\tilde{C}_p$ (S.59). The second term of Eq. (S.60) is also obtained by substituting (S.57) as

$$\langle \xi \circ v \rangle = \int_{-\infty}^{\infty}\left\langle \tilde{\xi}^* \circ \tilde{v}\right\rangle\frac{d\omega}{2\pi}$$

$$= \int_{-\infty}^{\infty}\left\langle \frac{\tilde{\xi}^* \circ k\tilde{v}_m + \tilde{\xi}^* \circ (k\tilde{R}_m + i\omega)\tilde{\xi} + \tilde{\xi}^* \circ i\omega\tilde{F}_p}{k(1+\gamma\tilde{R}_m) + i\omega\gamma}\right\rangle\frac{d\omega}{2\pi}$$

$$= \int_{-\infty}^{\infty} 2k_BT\gamma\frac{k\tilde{R}_m + i\omega}{k(1+\gamma\tilde{R}_m) + i\omega\gamma}d\omega$$

$$= \gamma\int_{-\infty}^{\infty} 2k_BTR_p'\frac{d\omega}{2\pi},$$

(S.63)

where we used the relationship $\left\langle\tilde{\xi}^* \circ \tilde{v}_m\right\rangle = \left\langle\tilde{\xi}^*\right\rangle = 0$ and $\left\langle\tilde{\xi}^2\right\rangle = 2k_BT\gamma$ at the third line and the definition of $\tilde{R}_p$ [Eq. (S.36)] at the fourth line. Because the imaginary part of $\tilde{R}_p$ is an odd function, the integration vanishes.

By substituting Eqs. (S.62) and (S.63) into (S.60), we obtained the total dissipation rate, $J_x$, as

$$J_x = \gamma\bar{v}_p^2 + \gamma\int_{-\infty}^{\infty}\left[\tilde{C}_p - 2k_BTR_p'\right]\frac{d\omega}{2\pi}. \quad (S.64)$$

This is exactly the same as the Harada-Sasa equality (S.14), indicating that the Harada-Sasa equality holds in our kinesin model. Thus, the total dissipation rate can be analytically obtained by calculating the righthand side of Eq. (S.64).

## IV. SUPPLEMENTARY DISCUSSIONS

### A. FRR of the kinesin motor

Here we derive the FRR of the kinesin motor at $\omega\to\infty$. The velocity fluctuation and the response function for the kinesin motor at $\omega\to\infty$ are calculated from Eqs. (S.52) and (S.30) as

$$\lim_{\omega\to\infty}\tilde{C}_m(\omega) = \frac{d^2k_c(k_f + k_b)}{k_a}$$

$$= \frac{d\cdot k_c}{k_a}(k_fd + k_bd) \quad (S.65)$$

$$\lim_{\omega\to\infty} 2k_BT\tilde{R}_m'(\omega) = 2k_BT\frac{d(k_fd_f - k_bd_b)k_c}{k_ak_BT}$$

$$= \frac{d\cdot k_c}{k_a}(2k_fd_f - 2k_bd_b). \quad (S.66)$$

Since these two formulas are different, the FRR is not generally satisfied at frequencies $\omega\to\infty$ in our kinesin model. So, we define $\Delta$ as the measure of the FRR violation at the

high frequency limit:

$$\Delta \equiv \lim_{\omega \to \infty}\left[\tilde{C}_m(\omega) - 2k_BT\tilde{R}'_m(\omega)\right]$$
$$= \frac{d \cdot k_c}{k_a}\left[(k_f d + k_b d) - (2k_f d_f - 2k_b d_b)\right] \quad (S.67)$$
$$= \frac{d \cdot k_c}{k_a}\left[k_f(d - 2d_f) + k_b(d + 2d_b)\right].$$

From this, we can obtain several corollaries: if the discrepancy, $\Delta$, is not equal to zero, the nonequilibrium dissipation from the kinesin motor diverges, and if $\Delta$ is negative, the nonequilibrium dissipation from the probe is also negative. These unrealistic corollaries do not seem to satisfy thermodynamic consistency. The reason for the apparent inconsistency may be due to ignoring the effect of the finite size of the kinesin molecule, as described in the next section.

It should be noted that this finite discrepancy exists even when the local detailed balance conditions and three reverse transition paths (Fig. 3a *dashed arrows* in the main text) are incorporated into the model; in both cases, the discrepancy equals zero when all characteristic distances are half the stepsize and/or the dissipation during each transition is negligibly small (Ref. [12] and Wang's *personal communications*).

$\Delta$ is mainly derived from the relationship between the stepsize and the characteristic distances. From realistic conditions for kinesin, the characteristic distances, $d_f$ and $d_b$, are not equal to half the stepsize, $d$. On the other hand, using 2D potential descriptions with an *observable* mechanical axis and a *hidden* chemical axis, Harada and Nakagawa proposed that the discrepancy between the characteristic distance and the stepsize reflects the irreversibility of molecular motors [13]. Therefore, the finite $\Delta$ may be related to the irreversibility of the system, although its physical meaning in the coarse grained Markov chain model has not been established.

## B. FRR of kinesin model considering head size

Our simplified Markov model of the kinesin motor neglected the motor size and the thermal fluctuation that directly perturbs the motor. As mentioned in the main text, the kinesin head has finite (~5 nm) size; the viscous drag of the motor itself should attenuate violation of the FRR of the motor at the high frequency limit even if there is no probe attached. Here, in order to evaluate the FRR of the kinesin motor including the finite size and the thermal fluctuation effects, we include these neglected effects into the simplified Markov model. This is done by using the same approach as used for the probe's dynamics (Fig. 3b in the main text), but by replacing the probe with the head of the motor; in the Langevin equation for the probe (S.31), we replace the probe's viscous drag, $\gamma$, and the stalk's spring constant, $k$, with the kinesin's head drag, $\gamma_k$, and spring constant, $k_k$, for the connection between the floating (tethered) head and the head bound to microtubule. The dynamics for the kinesin motor with finite size is given as

$$\gamma_k v_k = k_k(x_m - x_k) + F_m + \xi_k, \quad (S.68)$$

where $x_k$ and $v_k$ are the kinesin motor's position and velocity, respectively, and $\xi_k$ is the thermal fluctuation for the motor given by $\langle\xi_k\rangle = 0$ and $\langle\xi_k(t)\xi_k(t')\rangle = 2k_BT\gamma_k\delta(t-t')$. Similar to the derivation in the preceding section, we obtain the velocity fluctuation and the response function for the kinesin motor:

$$\tilde{C}_k(\omega) = \frac{k_k^2\tilde{C}_m(\omega) + 2k_BT\gamma_k\left[\omega^2(1+k_kR''_m)^2 + k_k^2R'^2_m\right]}{k_k^2(1+\gamma_kR'_m)^2 + \omega^2\gamma_k^2(1+k_kR''_m)^2}$$
$$\sim \tilde{C}_m(\omega) \quad (\gamma_kR'_m \ll 1, \omega \ll k_k/\gamma_k) \quad (S.69)$$
$$\sim 2k_BT/\gamma_k \quad (\omega \gg k_k/\gamma_k)$$

$$2k_BT\tilde{R}'_k = 2k_BT\frac{k_k^2R'_m(1+\gamma_kR'_m) + \omega^2\gamma_k(1+k_kR''_m)^2}{k_k^2(1+\gamma_kR'_m)^2 + \omega^2\gamma_k^2(1+k_kR''_m)^2}$$
$$\sim 2k_BTR'_m \quad (\gamma_kR'_m \ll 1, \omega \ll k_k/\gamma_k) \quad (S.70)$$
$$\sim 2k_BT/\gamma_k \quad (\omega \gg k_k/\gamma_k).$$

The second lines for each equation indicate that both the velocity fluctuation and the response for the kinesin motor approximate conditions that neglect head size when the viscous drag, $\gamma_k$, is sufficiently small ($\gamma_kR'_m \ll 1$) and the frequency is below the cutoff $k_k/\gamma_k$. On the other hand, both the velocity fluctuation and the response approach the value $2k_BT/\gamma_k$ at the high frequency limit, thus the FRR violation decays to zero over the cutoff frequency.

Because the head size is almost 1/100 times smaller than the probe, viscous drag on the head can be estimated as $\gamma_k \approx 3 \times 10^{-7}$ pN/nm·s. The floating head is tethered by a disordered 14 amino acids peptide chain called the "neck linker". The stiffness was estimated as $k_k \approx 1$ pN/nm [14]. Thus we can roughly estimate the cutoff frequency as hundreds of kHz. By integrating the difference between Eq. (S.69) and Eq. (S.70) with these parameters, the nonequilibrium dissipations from the head were estimated as ≈500 pN·nm/s (high ATP) and ≈5 pN·nm/s (low ATP). These values are much smaller than the ≈80% hidden dissipations (≈5000 pN·nm/s at high ATP and ≈1800 pN·nm/s at low ATP), suggesting that the origin of most of the hidden dissipation cannot be explained by dissipation only from the translational motion of the kinesin motor. Note that the neck linker is an entropic spring so that the stiffness becomes much higher when the linker is extended. However, even if the stiffness is a hundred times higher than the value estimated above, the hidden dissipation (nearly proportional to the stiffness) cannot be explained, at least at low ATP conditions.

---

## V. SUPPLEMENTARY REFERRENCES